\documentclass[]{interact}

\usepackage{epstopdf}
\usepackage[caption=false]{subfig}

\usepackage[numbers,sort&compress]{natbib}
\bibpunct[, ]{[}{]}{,}{n}{,}{,}
\renewcommand\bibfont{\fontsize{10}{12}\selectfont}
\makeatletter
\def\NAT@def@citea{\def\@citea{\NAT@separator}}
\makeatother

\theoremstyle{plain}

\theoremstyle{definition}

\theoremstyle{remark}

\begin{document}


\title{Multi-agent control of airplane wing stability \\under the flexural torsion flutter}

\author{
\name{Dmitry.~S. Shalymov\textsuperscript{a,b}, Oleg N. Granichin\textsuperscript{a,b},
Zeev Volkovich\textsuperscript{c} and Gerhard-Wilhelm Weber\textsuperscript{d}
\thanks{CONTACT D.~S. Shalymov. Email: dmitry.shalymov@gmail.com} 
}
\affil{
\textsuperscript{a}St. Petersburg State University, Universitetskaya nab. 7-9, Saint-Petersburg, 199034, Russia; 
\textsuperscript{b}IPME RAS, V.O., Bolshoj pr., 61, St. Petersburg, 199178, Russia;
\textsuperscript{c}Software Engineering Department, Ort Braude College, Rehov Snunit 51, POB 78, Karmiel 2161002, Israel;
\textsuperscript{d}Politechnika Poznanska, Wydzial Inzynierii Zarzadzania, ul. Jacka Rychlewskiego 2, 60-965 Poznan, Poland
}
}

\maketitle

\begin{abstract}
This paper proposes a novel method for prevention of the increasing oscillation of an aircraft wing under the flexural torsion flutter. The paper introduces the novel multi-agent method for control of an aircraft wing, assuming that the wing surface consists of controlled 'feathers' (agents).
Theoretical evaluation of the approach demonstrates its high ability to prevent flexural-torsional vibrations of an aircraft. Our model expands the possibilities for damping the wing oscillations, which potentially allows an increase in aircraft speed without misgiving of flutter.
The study shows that the main limitation is the time, during which the system is able to damp vibrations to a safe level and keep them. The relevance of this indicator is important because of the rather fast process of increasing wing oscillations during flutter. 
In this paper, we suggest a new method for controlling an aircraft wing, with the use of which it becomes theoretically possible to increase the maximum flight speed of an aircraft without flutter occurrence. A mathematical model of the bending-torsional vibrations of an airplane wing with controlled feathers on its surface is presented. Based on the Speed-Gradient method a new control laws are synthesized.
\end{abstract}


\section{INTRODUCTION}

In this paper, we investigate airplane wing stability under the flexural torsion flutter. This problem is strategically important for the aircraft industry.

In fact, after reaching a certain flight speed, small oscillations of the wing begin to rapidly and catastrophically increase until the wing breaks. The stability of an aircraft wing characterizes the wing's ability to maintain its integrity, when oscillations occur.

The design of the aircraft is resilient. During the flight, its elements can experience significant strain. For example, on a heavy transport aircraft, even in horizontal flight at a constant speed, the deflection of the wing end is measured in meters, cf. \cite{A1}. Deformations of the wing affect the size and distribution of the aerodynamic load. They can lead to a loss of structural stability, both static (e.g. wing divergence) and dynamic (e.g. flutter). 

Therefore, the issues of ensuring the required aerodynamic characteristics and wing stability in various flight modes of the aircraft are traditionally paid great attention to when designing. It should be noted that the elements of wing mechanization that are widespread in modern aircraft building (pre-flaps, ailerons, flaps, etc.) serve precisely this purpose; in particular, in the most critical take-off and landing modes. However, there are no ways to effectively counter wing flutter. 

Flutter is a phenomenon, which occurs after reaching a certain flight speed $V_{flat}$, when small wing oscillations begin to increase until the wing breaks. To avoid such consequences, currently one imposes a limit on the maximum flight speed of the aircraft, calculated as $(1.2-1.3)V_{max}=V_{flat}$. This significantly reduces the potential technical characteristics of the aircraft.

As a rule, an order to avoid the occurrence of flutter, lots of limitations are introduced on the maximum flight speed of the aircraft, which depends on the speed, at which the flutter $V_{flat}$ occurs. This practice significantly lowers the potential performance of the aircraft. By effectively avoiding flutter, it becomes possible to remove such restrictions and use the aircraft more efficiently. 

Until now, there are no effective methods for dealing with flutter. The designers strive to increase $V_{flat}$ as much as possible in order to raise the aircraft's high-speed ceiling. However, the existing approaches to increasing $V_{flat}$ are strongly limited by the features of specific aircraft models and their characteristics.

In this paper, in order to solve the described problem, we propose to cover an aircraft wing with lots of small-sized movable elements, which are capable of purposefully changing their orientation in the air flow. The movable elements allow creating of additional forces and moments, acting on the wing and preventing the growth of its deformations. 

\begin{figure}[thpb]
	\centering
	\includegraphics[scale=0.4]{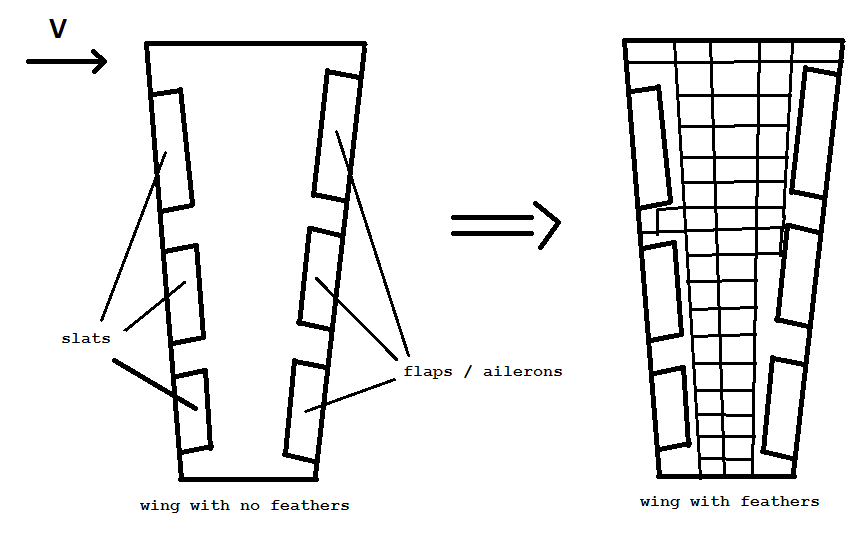}
	\caption{Wing with feathers}
	\label{fig:1p}
\end{figure}

Suppose that the entire free surface of the wing is covered with such controllable elements:  we call them feathers. The wing is covered with such feathers both above and below, and in the neutral position, they do not change the calculated wing profile, see Fig.\ref{fig:1p}. The last condition ensures that the wing dynamics does not deteriorate with passive feathers in the neutral position.

The novelty of our approach lies in the use of 'feathers' on the surface of the aircraft. In this way, the wing of the aircraft becomes a multi-agent system, which we control, based on methods from control theory. This approach is new and has not been used before in the problem of dealing with flutter.

The proposed approach should improve the existing situation due to the following: the controlled elements (feathers) can change their position in the air flow in such a way that they prevent the occurrence of oscillations and general deformation of the wing. Moreover, the feathers in their neutral position do not change the calculated wing profile. This ensures that the dynamics of the wing do not deteriorate even if they are turned off and are not used (which corresponds to passive feathers in neutral).

In this paper, we propose the laws of control of an airplane wing with 'feathers'. However, these laws are rather theoretical ones. In practice, a critical performance indicator is the time, needed for the system to damp and maintain vibrations to a safe level. The relevance of this indicator is due to the rather rapid process of increasing wing oscillations during flutter. In theory, we believe that the 'feather' changes its orientation in the air flow instantly. Unfortunately, in real physical systems, there is always a delay. The impact of the delay on our theoretical conclusions should be rather investigated in the further works.

The rest of the paper is structured in the following way: Section 2 provides overview of related work. Section 3 is devoted to study of the equation of system dynamics for small flexural-torsional vibrations of a wing with feathers on the surface.  Section 4 describes the synthesis of the dynamic system control law, based on the speed-gradient method. A control law is obtained in the form of feedback on deviation with constant coefficients. Section 5 introduces the transition to multi-agent control. Section 6 summarizes the paper.

\section{RELATED WORK}

The flight condition boundary between stable and self-sustaining motions is known as flutter speed and flutter boundary.  There are several types of flutter, which significantly vary depending on the way, in which the stability is lost with the any change in flight conditions like increase in dynamic pressure. An explosive flutter occurs after a small increase of speed from just below the flutter speed to slightly above the flutter speed. This results in highly divergent oscillations and wing break within a fraction of a second.

The moderate flutter occurs when loss of stability can be identified well below the flutter speed. In this case, the flutter speed can be predicted by extrapolating instabilities. The mild type flutter occurs when the system is stable but lowly damped significantly before the flutter speed.

Unlike traditional stable aircraft flight control systems, which aim to improve stability, improve handling, gust alleviation, control ride comfort etc., the approach which aims to stabilize an unstable system is called Active Flutter Suppression (AFS). The AFS system must stabilize an aeroelastic system that would otherwise be unstable in all flight conditions and aircraft maneuvers, covering all configuration and loading options as well as all flutter mechanisms. That is why AFS is important for effective solution of aeroelastic instability problems. AFS can lead to significant weight savings and more efficient and versatile airframes as described in particular in \cite{c6}. A broad overview of AFS research is rather presented in \cite{G1}. 
The ability to suppress flutter instability through actively controlled closed-loop action of control surfaces has a long history, see, for example \cite{c5}. Numerous studies in the field of AFS were carried out as early as the 1970s and 1980s, cf.  \cite{c9}. Various approaches to the synthesis of AFS systems control law, including methods of adaptive control and control with variable parameters, are considered also in \cite{c364,c366,c369,c370}.

The physics-based approach of AFS research of \cite{c371} investigates the laws of flutter control, based on searching the physical or mathematical structure of the flutter problem in order to identify the mechanisms responsible for the flutter instability and find the ways to suppress them.
The concepts of ‘active flexible wing’ or ‘active flexible airframe’ are considered in \cite{c57}; flight stability and controllability of rigid and flexible aircraft are considered in \cite{c15, c23, c27}; the influence of aeroelastic behavior on stability and controllability of flight using corrections to the derivatives of static aeroelastic stability was studied in \cite{c31}; the perspective of using active controls is presented in \cite{c134}.

The idea of using an active control system, according to which special mechanisms are located on the wing of the aircraft that control the surfaces of the airframe to achieve the desired dynamic behavior, has been considered and discussed from the first days of manned flights, cf.\cite{c15,c23}. Adaptive control is attractive in the case of flutter suppression due to many variations in plant characteristics and ability to respond to damage scenarios, cf. \cite{c417,c418,c420,c422,c423,c424}. In \cite{c403}, an attempt is made to order and reduce various laws of modern control theory used to suppress flutter.

Control problems of nonlinear systems, as well as topological aspects of active systems, are considered in \cite{c434, c437, c438, c439, c443}. A broad review of methods for the synthesis of control laws and modeling of aeroelastic systems is rather presented in \cite{c444}. The identically located accelerometers and forces method is used to create an effective viscous damping matrix for the equations of the system that would stabilize it \cite{c369}.
A similar approach to the active control of structures with guaranteed stability is presented in \cite{c370}. 
Control laws that modify the net stiffness and mass distribution are investigated in \cite{c366}.

Other aspects of the problem of the flutter suppression law are discussed in \cite{c425,c443}. The influence of control system hardware delays, which is very important in the case of flutter, is considered in \cite{c425}. A special approach to active control of systems with changing parameters 
and control of nonlinear aeroelastic systems is investigated 
in \cite{c434, c437, c438, c439}. Broader prospects for the use of AFS in practice appeared with the development of fast drives and actuators \cite{c86}. Overview of aeroservoelasticity is done in 
\cite{c152}.

Today it is possible to carry out high accuracy aerodynamic modeling of active controlled systems based on control laws, which were usually synthesized using 
low-order mathematical models \cite{c181, c186}. However, such models still pose a serious problem for developers and designers due to their large size.

The flutter behavior is investigated in numerous more recent works; cf. \cite{G3,G6}. A control system, applied to transonic flutter suppression, is presented in \cite{G5}. Robustness to measurement noises and time delay is demonstrated. Stochastic modeling, based on the first-order reliability method for flutter reliability analysis of an aircraft wing is proposed in \cite{G3}.
General aero-elastic system in a subsonic regime is investigated in \cite{G4}. As part of aeroelastic analysis a flutter speed for a more fuel efficient slender wing is identified in \cite{G2}. A systematic robust control design method for AFS is provided in \cite{G6}. The method is used to increase the flutter speed of a small unmanned aircraft.

AFS continues to be considered with caution and is prohibited from use on commercial and military aircraft. This is because a failure of the AFS system while the divergent flutter is unstable can lead to the failure of the airframe so quickly that the crew cannot react in any way, for example, by reducing the flight speed. 

Multi-agent systems (MAS) have numerous applications like civilian, security and military ones. Centralized quantitative and qualitative modeling, analysis, constraint satisfaction, maintenance and control seem to be too strict for these systems. On the other hand, the distributed and incremental reasoning on the systems seems to be more scalable, robust, and flexible. That is why an investigation of multi-agent control systems is popular nowadays and has been researched for decades, cf.  \cite{F1, Gr3, Fradkov1}. Multiagent approach is able to replace the general model of the interactions in a complex system with a set of local models. This can be effectively applied in distributed nonstationary systems such as dynamical networks of \cite{G3,F12}, where the problem of consensus is investigated. Agents can be modeled as integrators on segments, cf. \cite{F7, Fradkov1}. The problem of control goal achievement for such agents is closely related to the consensus problems of \cite{F12, Wiener21} and can be reduced to it in some cases as described in \cite{F7}. Consensus in multi-agent network systems from MaxEnt perspective is investigated in \cite{Wiener21}. 
For more details, see, for example, \cite{EVR1, EVR2}. 

Dynamical networks and multi-agent protocol for the airplane is investigated in \cite{Gran1,A5}. The wing is covered with controlled agents 'feathers' and multi-agent control is used for leveling the perturbing forces in the turbulence. Each 'feather' is controlled based on a feedback about pressure of its own and neighbor feathers.

\section{DYNAMICS EQUATIONS OF WING WITH FEATHERS}

A wing with feathers has never been considered before for the problem of flutter suppression. Therefore, in this section we provide description of our novel approach, based on the feathers. Let us consider the phenomenon of flexural-torsional flutter of the wing as described in \cite{A2,A3}. Let the plane be in a steady horizontal flight at a constant speed. We schematize the non-sweeping wing of the half-span l and feathers in neutral position with a cantilevered beam. This beam is loaded with a static distributed load on bending and torsion.
The elastic axis of the wing passes through the Stiffness Centers (SC's) of the sections and does not coincide with the line of Gravity Centers (GC's) of the sections. We assume the wing stiffness in the longitudinal and transverse directions of the wing plane to be very large. The assumption allows neglecting of these types of vibrations. We also neglect possible movements of the SC and GC along the sections during the flight.

\begin{figure}[thpb]
	\centering
	\includegraphics[scale=0.3]{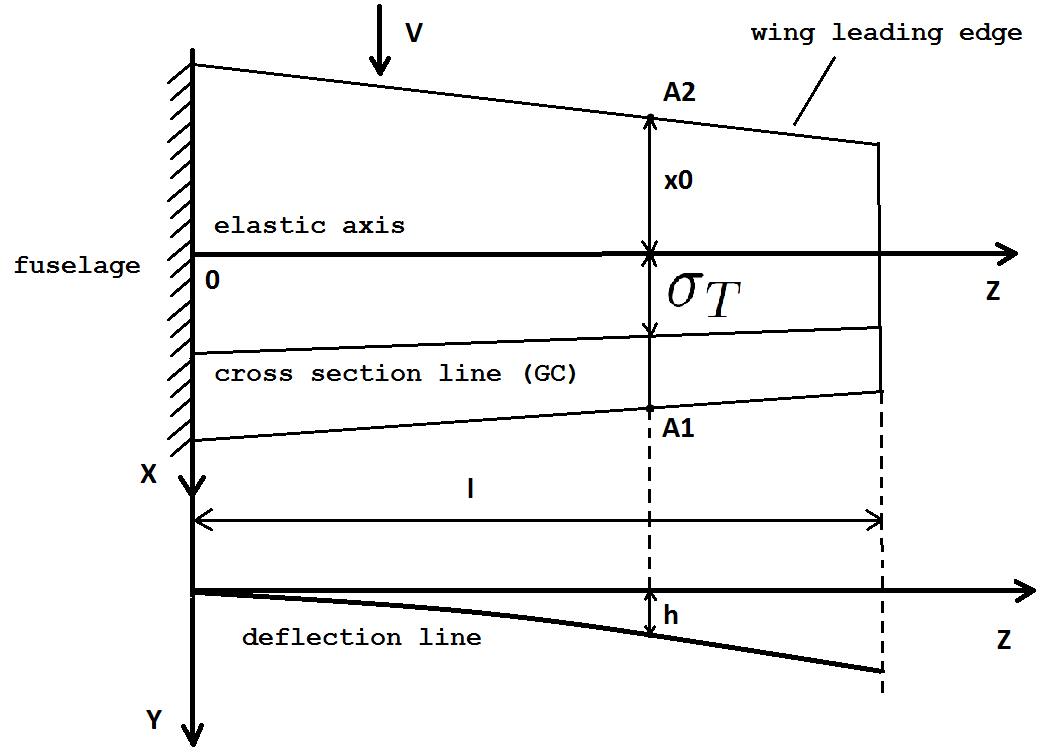}
	\caption{Stiffness Centers and Gravity Centers lines on a wing}
	\label{fig:2p}
\end{figure}

Figure \ref{fig:2p} illustrates the following: 
\begin{itemize}
	\item $O$ is the stiffness center of the wing section on the fuselage;
	\item $X$-axis is directed on a free stream;
	\item $Z$-axis is directed along the elastic axis of the undeformed wing;
	\item $Y$-axis complements the coordinate system to the right;
	\item $x_{0}$ - is the distance from the leading edge of the wing to the SC section;
	\item $\sigma_{T}$ - is the distance between SC and GC.
\end{itemize}

We note that typically the GC is located behind the SC section.

\begin{figure}[thpb]
	\centering
	\includegraphics[scale=0.5]{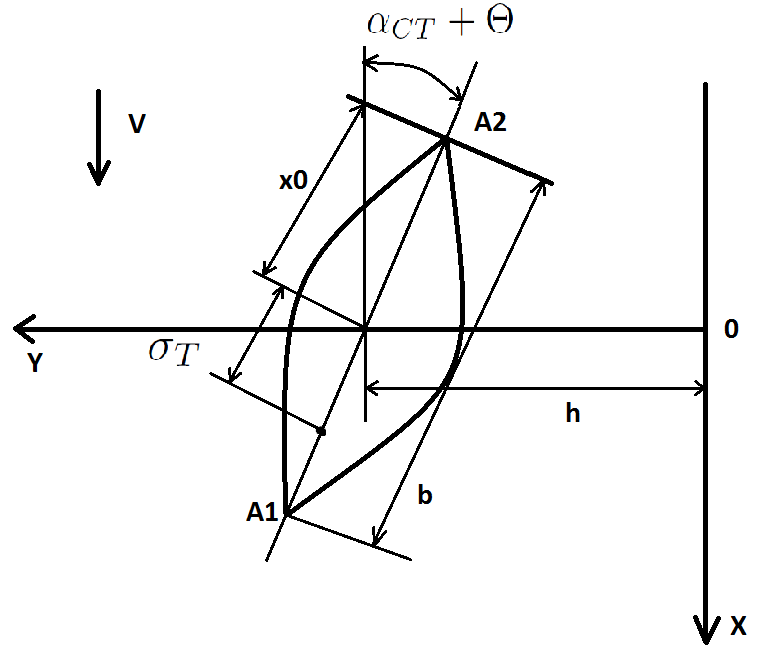}
	\caption{Wing cross section}
	\label{fig:3p}
\end{figure}

Figure \ref{fig:3p} illustrates the following: 
$b$ is the wing section chord;
$h$ is the transverse deflection of the section $A_{1}A_{2}$;
$\alpha_{CT}$ is the angle of attack in the section of the un-deformed wing;
$\Theta$ is the angle of twisting of the section due to torsional deformations of the wing.

The corresponding equations of the elastic line of the beam are of the form, cf. \cite{c5}:

\begin{equation}\label{eq1}
\begin{cases}
\frac{\partial^{2}}{\partial z^{2}}\left( EJ\frac{\partial^{2}y}{\partial z^{2}}\right) = q^{0},\\
\frac{\partial}{\partial z}\left( GJ_{K}\frac{\partial\Theta}{\partial z}\right) = m^{0},
\end{cases}
\end{equation}
where 
\begin{itemize}
	\item $y$ is the deflection of the stiffness axis in the current wing section;
	\item $\Theta$ is the wing twist angle, which is considered positive if it increases the angle of attack in the section;
	\item $EJ, GJ_{k}$ are the wing stiffness in bending and torsion, respectively;
	\item $q^{0}$ and $m^{0}$ are the linear force and moment relative to the stiffness axis acting on the wing.
\end{itemize}

The functions of Eq. (\ref{eq1}) do not depend on time, since the wing is in a steady (stationary) state; i.e. solutions of Eq. (\ref{eq1}) can depend only upon z:
\begin{equation}\label{eq2}
y=y^{0}(z),~\Theta=\Theta^{0}(z).
\end{equation}

These solutions must satisfy the boundary conditions at the ends of the wing:
\begin{equation}\label{eq3}
\begin{cases}
\left. y\right\rvert_{z=0}=0; ~~ \left. \frac{\partial y}{\partial z}\right\rvert_{z=0}=0~~ (\textnormal{tight fuselage fit});\\

\left. EJ\frac{\partial^{2}y}{\partial z^{2}}\right\rvert_{z=l}=0;~~ 
\left. \frac{\partial}{\partial z}\left(EJ\frac{\partial^{2}y}{\partial z^{2}}\right) \right\rvert_{z=l}=0 \\
~~ (\textnormal{moment and shear force at the free end}),\\

\left. \Theta\right\rvert_{z=0}=0; ~~ \left. GJ_{K}\frac{\partial \Theta}{\partial z}\right\rvert_{z=l}=0 ~~ \\
(\textnormal{angle of rotation of the terminated end and moment at the free end}).
\end{cases}
\end{equation}

Now, suppose that for some random reason (sudden aileron movement, air hole, gust of wind, etc.), the wing deviated from its stationary position (see Eq. ((\ref{eq2}))). After termination of the reason of the deviation, the wing under the action of elastic forces will move to its position of equilibrium. If the energy dissipation is not very big then the aperiodic process will not occur; however, the wing oscillations will occur. We assume that these fluctuations are initially small enough and do not significantly affect the dynamics of the aircraft.

According to \cite{A2,A3}, small bending-torsional oscillations of the wing, near its equilibrium position (see Eq. (2, 3)) in the laminar flow, are described by the following equations:
\begin{equation}\label{eq4}
\begin{cases}
\frac{\partial^{2}}{\partial z^{2}}\left( EJ\frac{\partial^{2}y_{1}}{\partial z^{2}}\right) + m\frac{\partial^{2}y_{1}}{\partial t^{2}} - m\sigma_{T}\frac{\partial^{2}\Theta_{1}}{\partial t^{2}}=q_{a}, \\

\frac{\partial}{\partial z}\left( GJ_{K}\frac{\partial\Theta_{1}}{\partial z}\right) + m\sigma_{T}\frac{\partial^{2}
y_{1}}{\partial t^{2}} - J_{m}\frac{\partial^{2}\Theta_{1}}{\partial t^{2}}=m_{a}, 
\end{cases}
\end{equation}

where 
\begin{itemize}
	\item $y_{1} $ and $\theta_{1}$ are the additional deflection and angle of twisting of the wing relative to the stationary state (see Eq. (\ref{eq2}),(\ref{eq3})), due to fluctuations;
	\item $m$ is the linear mass of the wing;
	\item $J_{m}$ is the linear mass moment of inertia of the wing relative to its stiffness axis;
	\item $q_ {a}$ and $m_ {a}$ are the linear aerodynamic force of the wing and the linear moment of the aerodynamic force relative to the stiffness axis, due to wing vibrations.
\end{itemize}

Solutions of the system (\ref{eq4}) must satisfy boundary conditions similar to (\ref{eq3}). We represent the right-hand sides of Eq. (\ref{eq4}) in the form:
\begin{equation}\label{eq5}
q_{a}=\Delta q_{a} + q_{u}, ~~ m_{a} = \Delta m_{a} + m_{u},
\end{equation}

where 
\begin{itemize}
	\item $\Delta q_ {a}$ and $\Delta m_ {a}$ are the linear aerodynamic force and moment relative to the stiffness axis respectively, arising due to wing oscillations in the neutral position of the feathers;
	\item $q_{u}$ and $m_{u}$ - the linear aerodynamic force and moment created by changing the orientation of the feathers.
\end{itemize}

Following the flutter equations (cf. \cite{A2}, p. 176, Eq. (35)) and taking into account linear aerodynamic forces and moment $q_{u}$ and $m_{u}$, we rewrite Eq. (\ref{eq4}) as

\begin{equation}\label{sist6}
\begin{cases}
\frac{\partial^{2}}{\partial z^{2}}(EJ\frac{\partial^{2}y_{1}}{\partial z^{2}})
+m\frac{\partial^{2}y_{1}}{\partial t^{2}}
-m\sigma_{T}\frac{\partial^{2}\Theta_{1}}{\partial t^{2}} \\
-C_{y}^{\alpha}\left[\Theta_{1}+(\frac{3}{4}b-x_{0})\frac{1}{V}\frac{\partial\Theta_{1}}{\partial t} -\frac{1}{V}\frac{\partial y_{1}}{\partial t}\right]\rho bV^{2}=q_{u}\\

\frac{\partial}{\partial z}(GJ_{K}\frac{\partial\Theta_{1}}{\partial z}) + m\sigma_{T}\frac{\partial^{2}y_{1}}{\partial t^{2}} - J_{m}\frac{\partial^{2}\Theta_{1}}{\partial t^{2}} 
- \frac{\pi}{16}\frac{b^2}{V}\frac{\partial\Theta_{1}}{\partial t} \rho bV^{2} \\
+\left\{
+C_{y}^{\alpha}(x_{0}-\frac{b}{4})\left[\Theta_{1}+(\frac{3}{4}b-x_{0})\frac{1}{V}\frac{\partial\Theta_{1}}{\partial t} -\frac{1}{V}\frac{\partial y_{1}}{\partial t}\right]
\right\}\rho bV^{2} = m_{u}, \\

y_{1} = \frac{\partial y_{1}}{\partial z}=\Theta_{1} = 0,~~z=0,\\

\frac{\partial^{2}y_{1}}{\partial z^{2}} = \frac{\partial^{3}y_{1}}{\partial z^{3}} = \frac{\partial\Theta_{1}}{\partial z}=0,~~z=l,

\end{cases}
\end{equation}

where 
\begin{itemize}
	\item $C_{y}^{\alpha}=\frac{\partial C_{y}}{\partial \alpha}$; $C_{y}$ is the wing lift coefficient;
	\item $C_{y}^{\alpha}$ consider constant along the span;
	\item $C_{y}=C_{y}^{\alpha}(\alpha-\alpha_{0});$
	\item $\alpha=\alpha_{CT}+\Theta^{0}+\Theta_{1}$ is the instant value of the angle of attack when the wing moves;
	\item $\alpha_{0}$ is the value of the angle of attack at which $C_{y}=0$;
	\item $\rho$ is the air density.
\end{itemize}

As it can be seen from Eq. (\ref{sist6}), the bending and torsional vibrations of the wing are interdependent, which is one of the necessary conditions for the occurrence of flutter. 
It is also known from \cite{c5} that with increasing speed V, the frequencies of the bending and torsional oscillations of the wing approach each other, and for $V = V_{flat}$ they coincide. Moreover, there is a phase shift between these oscillations, cf.  \cite{A2,A3}. The last is also a necessary condition for the occurrence of flutter.

It is important to emphasize that for $V = V_{flat}$ the amplitude of the wing oscillations is still small but remains constant. It means that the oscillations themselves are no longer self-damped, as it was the case with $V < V_{flat}$. The problem arises when $V > V_{flat}$, i.e. when the slightest deformations of the wing begin to grow catastrophically quickly. Therefore, in terms of Eq. (\ref{sist6}), to prevent flutter, it is necessary to form qu and mu so that the wing oscillation energy is always limited by a certain value. This value is safe from the point of view of controllability, stability and integrity of the aircraft structure. In other words, when speeds $V > V_{flat}$, for the full energy of the cantilevered beam, cf \cite{A2}, the condition must be:

\begin{multline}\label{eq7}
E = E_{kinet} + E_{poten} = 
\frac{1}{2}\int_{0}^{l}{m\left(\frac{\partial y_{1}}{\partial t} \right)^{2}dz}
+ \frac{1}{2}\int_{0}^{l}{J_{m}\left(\frac{\partial \Theta_{1}}{\partial t} \right)^{2}dz} 
-\int_{0}^{l}{m\sigma_{T}\frac{\partial y_{1}}{\partial t}\frac{\partial \Theta_{1}}{\partial t}dz} + \\ 
\frac{1}{2}\int_{0}^{l}{EJ\left(\frac{\partial^{2}y_{1}}{\partial z^{2}} \right)^{2}dz} 
+\frac{1}{2}\int_{0}^{l}{GJ_{K}\left(\frac{\partial \Theta_{1}}{\partial z} \right)^{2}dz} \leq E_{*}.
\end{multline}

A stricter requirement is to keep the system (\ref{sist6}) in a given small neighborhood of the solution of Eq. (\ref{eq2}):
\begin{equation}\label{eq8}
\left\|\bar{x}\right\| \leq \epsilon,
\end{equation}
where 
\begin{equation}\notag
\bar{x} = 
\left( y_{1}, \frac{\partial y_{1}}{\partial t}, \Theta_{1}, \frac{\partial \Theta_{1}}{\partial t} \right)
\end{equation}
and 

$~\left\|\bar{x}\right\|$ is vector norm $\bar{x}$.

Power $q_u$ and moment mu impacts on the wing are formed due to the high-speed pressure and therefore should depend on the velocity of the flow of V, the position of the feathers on the wing, their orientation as well as other factors, associated with the adopted aerodynamic calculation scheme.

Let’s assume that the feathers are absolutely rigid structural elements, and that a change in the orientation of the feather does not have a significant effect on the airflow around the remaining feathers and the wing as a whole keeping it laminar. Then, we receive:
\begin{equation}\label{eq9}
q_{u} = \sum_{i}^{n(z)}{q_{u_{i}}}, ~~ m_{u} = \sum_{i}^{n(z)}{m_{u_{i}}},
\end{equation}
where $q_{u_{i}}$ and $m_{u_{i}}$ are the additional linear forces and the moment from the i-th feather. Summation is carried out over all the feathers, covering the boundary of the section. The number of such feathers depends on $z$ and is denoted as $n(z)$.

Suppose that a feather can deviate from its neutral position by rotating around an axis, passing through its leading edge and parallel to the Z-axis. 
Moreover, for the feathers on the upper surface of the wing, the rotation angle is measured from the tangent to the profile at the point of rotation 
of the feather $\beta_{i}\in[0,\beta^{-}],~\beta^{-}<0,$ and on the lower surface of the wing is measured as $\beta_{i}\in[0,\beta^{+}],~\beta^{+}>0,$ 
corresponds to the neutral position of the wing, and $\beta_{i}$ is considered positive if it locally increases the angle of attack.

\begin{figure}[thpb]
	\centering
	\includegraphics[scale=0.38]{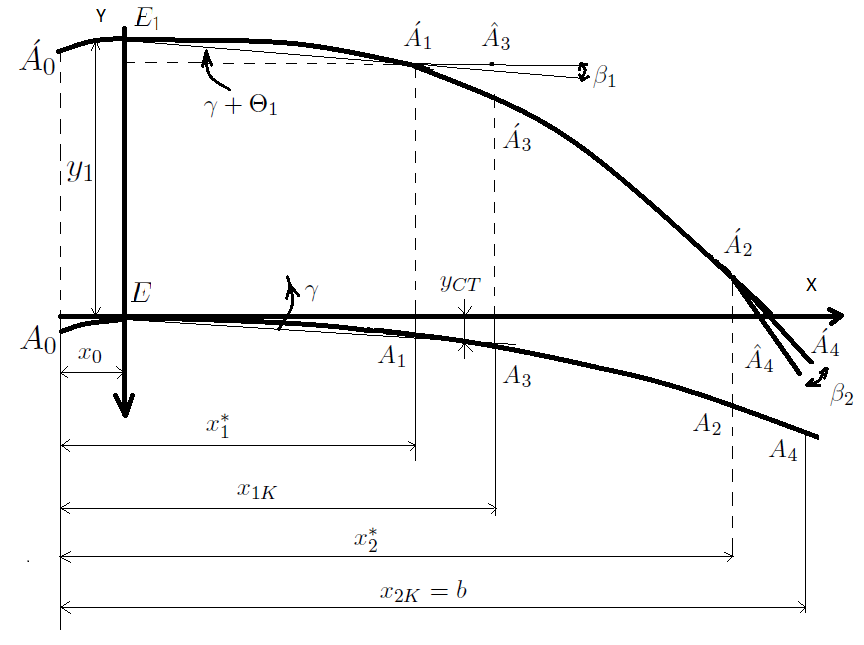}
	\caption{Wing profiles in static position and during vibrations}
	\label{fig:4p}
\end{figure}

In Fig. \ref{fig:4p}, we use the following notation:
\begin{itemize}
	\item $A_{0}~A_{1}~A_{2}~A_{4}$ are wing profiles (considered thin) in a static position (before vibrations);

	\item $\acute{A}_{0}~\acute{A}_{1}~\acute{A}_{2}~\acute{A}_{4}$ are wing profiles during oscillations; 

	\item $E$ and $E_{1}$ are SC's of the wing section in the static position and during vibrations, respectively;

	\item $X$-axis corresponds to the speed of the main stream;

	\item $Y$-axis is perpendicular to it and to the axis of rigidity of the undeformed wing;

	\item $A_{1}~A_{3}~\acute{A}_{1}~\acute{A}_{3}$ are the front and back edges of the feather 1 in the neutral position on the corresponding profiles;

	\item $A_{2}~A_{4}~\acute{A}_{2}~\acute{A}_{4}$ are the front and back edges of the feather 2 (analogue of the aileron) in the neutral position on relevant profiles;

	\item $x_{1}^{*}$ and $x_{1K}$ are the distances from the leading and trailing edges of the feather 1 to the leading edge of the wing;

	\item $x_{2}^{*}$ and $x_{2K}$ are similar parameters for the feather 2;

	\item $EE_{1}$ is deflection of the wing;

	\item $\Theta_{1}$ is the angle of twisting of the wing near the point $E_{1}$;

	\item $\beta_{1}<0$ is the angle of deviation of the feather 1 from the neutral position;

	\item $\beta_{2}>0$ is the angle of deviation of the feather 2 from the neutral position;

	\item $\hat{A}_{3}$ and $\hat{A}_{4}$ are the trailing edges of the feathers 1 and 2, respectively, after their deviations.

\end{itemize}

Parameters $\Theta_{1}$, wing deflection $y_{CT}$ are ordinates in a static position, $\beta_{i}$ are considered small.

A similar figure for the aileron is presented in (cf. \cite{A1} p. 143, Fig. 41).

Following the technique suggested by (cf. \cite{A2}, pp.143-146), it can be shown that the influence of the i-th feather on the wing is generally calculated using the following formulae:

\begin{equation}\label{eq10}
\begin{cases}
q_{u_{i}} = A_{i}V^{2}\beta_{i} + B_{i}V\dot{\beta}_{i}, \\
m_{u_{i}} = C_{i}V^{2}\beta_{i} + D_{i}V\dot{\beta}_{i},
\end{cases}
\end{equation}
where
$A_{i}=C_{y}^{\alpha}G_{i}\rho b^{2},~$
$B_{i}=C_{y}^{\alpha}H_{i}\rho b^{3},$

$C_{i}=-\left[ I_{i} + C_{y}^{\alpha}(\frac{x_{0}}{b}-\frac{1}{4})G_{i}\right]\rho b^{2},$

$D_{i}=-\left[ J_{i} + C_{y}^{\alpha}(\frac{x_{0}}{b}-\frac{1}{4})H_{i}\right]\rho b^{3},$

$G_{i}=\frac{1}{\pi}\left[ (\psi_{ik}-\psi_{i}^{*}) - (\sin{\psi_{ik}}-\sin{\psi_{i}^{*}})\right],$

\begin{multline}\notag
H_{i}=\frac{1}{2\pi}\left(
\cos{\psi_{i}^{*}}(\psi_{ik}-\psi_{i}^{*}) - (\sin{\psi_{ik}}-\sin{\psi_{i}^{*}})\right) 
-\cos{\psi_{i}^{*}}(\sin{\psi_{ik}}-\sin{\psi_{i}^{*}}) \\
+\frac{1}{2}\left(
(\psi_{ik}-\psi_{i}^{*}) +\frac{1}{2}(\sin{2\psi_{ik}}-\sin{2\psi_{i}^{*}})
\right);
\end{multline}

$I_{i}=\frac{1}{8}\left[ 2(\sin{\psi_{ik}}-\sin{\psi_{i}^{*}}) + (\sin{2\psi_{ik}}-\sin{2\psi_{i}^{*}})\right],$

\begin{multline}\notag
J_{i}=-\frac{1}{16}\left(
-2\cos{\psi_{i}^{*}}(\sin{\psi_{ik}}-\sin{\psi_{i}^{*}}) + (\psi_{ik}-\psi_{i}^{*})
\right)
-\frac{1}{16}\left(
(\frac{1}{2}-\cos{\psi_{i}^{*}})(\sin{2\psi_{ik}}-\sin{2\psi_{i}^{*}})
\right)
\\
-\frac{1}{16}\left( 
(\sin{\psi_{ik}}-\sin{\psi_{i}^{*}})
+\frac{1}{3}\left(
\sin{3\psi_{ik}}-\sin{3\psi_{i}^{*}}
\right)
\right);
\end{multline}

$x_{i}^{*}=\frac{b}{2}(1-\cos{\psi_{i}^{*}}),~~$
$x_{ik}=\frac{b}{2}(1-\cos{\psi_{ik}}),$

$\psi_{i}^{*}\in[0,\pi],~~$
$\psi_{ik}\in[0,\pi].$

Let us get back to Eq. (\ref{sist6}).
In \cite{A2,A3}, its solution near the flutter is represented as
\begin{equation}\label{eq11}
\begin{cases}
y_{1}(z,t)=q(t)f(z),\\
\Theta_{1}(z,t)=r(t)\phi(z).
\end{cases}
\end{equation}

Moreover, functions $f(z)$ and $\phi(z)$ are functions of vibration modes, which must satisfy the boundary conditions:
at $z=0,~f=0;~f'=0;~\phi=0;$
at $z=l,~f''=0;~f'''=0;~\phi'=0.$

Here for the sake of simplicity, we take $f'=\frac{\partial f}{\partial z}$,
$f'''=\frac{\partial^{3} f}{\partial z^{3}}.$

We substitute Eq. (\ref{eq11}) into Eq. (\ref{sist6}) and multiply the first equation by $f$ and the second equation by $\phi$ and then integrate from 0 to l. 
After simple transformations using Eq. (\ref{eq9}) and Eq. (\ref{eq10}), we obtain the following system
\begin{equation}\label{eq12}
\begin{cases}
a_{11}\ddot{q} + a_{12}\dot{q} + a_{13}q + b_{11}\ddot{r} + b_{12}\dot{r}+b_{13}r = Q(\beta,\dot{\beta}), \\
a_{21}\ddot{q} + a_{22}\dot{q} + b_{21}\ddot{r} + b_{22}\dot{r}+b_{23}r = M(\beta,\dot{\beta}),
\end{cases}
\end{equation}
where

$a_{11}=\int_{0}^{l}{mf^{2}dz},$

$a_{12}=C_{y}^{\alpha}\rho V\int_{0}^{l}{bfdz},$

$a_{13}=\int_{0}^{l}{\frac{d^{2}\left(EJf''\right)}{d{z}^{2}}fdz},$

$b_{11}=-\int_{0}^{l}{m\sigma_{T}f\phi dz},$

$b_{12}=-C_{y}^{\alpha}\rho V\int_{0}^{l}{\left(\frac{3}{4}b-x_{0}\right)bf\phi dz},$

$b_{13}=-C_{y}^{\alpha}\rho V^{2}\int_{0}^{l}{bf\phi dz},$

$a_{21}=\int_{0}^{l}{m\sigma_{T}f\phi dz}=-b_{11},$

$a_{22}=-C_{y}^{\alpha}\rho V\int_{0}^{l}{\left(x_{0}-\frac{b}{4}\right)bf\phi dz},$

$b_{21} = -\int_{0}^{l}{J_{m}\phi^{2}dz},$

$b_{22}=-\frac{\pi}{16}\rho V\int_{0}^{l}{b^{3}\phi^{2}dz} + 
C_{y}^{\alpha}\rho V\int_{0}^{l}{b(x_{0}-\frac{b}{4})\left(\frac{3}{4}b-x_{0}\right)\phi^{2} dz},$

$b_{23}=b_{23}^{(1)} + b_{23}^{(2)} = C_{y}^{\alpha}\rho V^{2}\int_{0}^{l}{b(x_{0}-\frac{b}{4})\phi^{2}dz} + 
\int_{0}^{l}{\frac{d\left(GJ_{k}\phi' \right)}{dz}\phi dz},$

$Q(\beta,\dot{\beta}) = \sum_{i=1}^{N}{\left(\bar{A}_{i}V^{2}\beta_{i} + \bar{B}_{i}V\dot{\beta}_{i}\right)},$

$\bar{A}_{i}=\int_{0}^{l}{A_{i}f dz},~~$
$\bar{B}_{i}=\int_{0}^{l}{B_{i}f dz},$

$M(\beta,\dot{\beta}) = \sum_{i=1}^{N}{\left( \bar{C}_{i}V^{2}\beta_{i} + \bar{D}_{i}V\dot{\beta}_{i}\right)},$

$\bar{C}_{i}=\int_{0}^{l}{C_{i}\phi dz},~~$
$\bar{D}_{i}=\int_{0}^{l}{D_{i}\phi dz},$

$\beta=col\left\{\beta_{i},~i={1,N} \right\},~~$
$\dot{\beta}=col\left\{\dot{\beta}_{i},~i={1,N} \right\},$
where $N$ is the total number of feathers on the wing surface.

Given two functions $f$ and $\phi$ and the known distributions of the mass and stiffness parameters of the wing (we consider them independent of t), 
the coefficients $a_{ij}$ and $b_{ij}, ~ i, j = 1,2,3$  can be calculated and will be constants. 
The accuracy of the further results depends on the choice of $f$ and $\phi$. 
Without going into details, we note that with good accuracy these functions can be calculated, for example, by the method of successive approximations.
We complement Eq. (\ref{eq12}) with equations for controls

\begin{equation}\label{eq13}
\dot{\beta} = u,
\end{equation}
where
$u = col\left\{u_{i},~~i={1,N} \right\};~~$
$\beta_{i}\in[0,\beta^{+}], ~~ \beta^{+} > 0, ~~ i \in \bar{1,n^{+}},~$
where $n^{+}$ - is the total number of feathers on the lower surface of the wing;

$\beta_{i}\in[\beta^{-},0], ~~ \beta^{-} < 0, ~~ i \in \bar{n^{+}+1,N},~$
where $n^{-}=N-n^{+}$ - total number of feathers on the upper surface of the wing.

Now, we use the following notation:
\begin{equation}\label{eq14}
x=col\left\{q, \dot{q}, r, \dot{r} \right\} = col\left\{x_{i},~~i=\bar{1,4} \right\}.
\end{equation}

Then, we substitute Eq. (\ref{eq14}) into Eq. (\ref{eq12}) and reduce this system to the normal Cauchy form. After that, after combining Eq. (\ref{eq12}) and Eq. (\ref{eq13}), we get

\begin{equation}\label{eq15}
\begin{cases}
\dot{x}_{1}=x_{2},\\
\dot{x}_{2}=\sum_{k=1}^{4}{C_{1k}x_{k}} + F_{1}(\beta,u),\\
\dot{x}_{3}=x_{4},\\
\dot{x}_{4}=\sum_{k=1}^{4}{C_{2k}x_{k}} + F_{2}(\beta,u),\\
\dot{\beta}=u,
\end{cases}
\end{equation}
where
$F_{1}(\beta,u)=d_{11}Q+d_{12}M=\sum_{i=1}^{N}{\left(R_{1i}\beta_{i}+s_{1i}u_{i} \right)},$

$F_{2}(\beta,u)=d_{21}Q+d_{22}M=\sum_{i=1}^{N}{\left(R_{2i}\beta_{i}+s_{2i}u_{i} \right)},$

$R_{1i}=V^{2}\left( \bar{A}_{i}d_{11} + \bar{C}_{i}d_{12}\right),$

$s_{1i}=V\left( \bar{B}_{i}d_{11} + \bar{D}_{i}d_{12}\right),$

$d_{11}=\left[a_{11}\left(1-\frac{a_{21}b_{11}}{a_{11}b_{21}} \right) \right]^{-1},$

$d_{12}=-d_{11}{b_{11}} / {b_{12}},$

$R_{2i}=V^{2}\left( \bar{A}_{i}d_{21} + \bar{C}_{i}d_{22}\right),$

$s_{2i}=V\left( \bar{B}_{i}d_{21} + \bar{D}_{i}d_{22}\right),$

$d_{21}=-a_{21}\frac{d_{11}}{b_{21}},$

$d_{22}=\left(1 - a_{21}d_{12} \right) / b_{21},$

$C_{11}=-d_{11}a_{13},$

$C_{12}=-d_{11}(a_{12}-b_{11}\frac{a_{22}}{b_{21}}),$

$C_{13}=-d_{11}(b_{13}-b_{11}\frac{b_{23}}{b_{21}}),$

$C_{14}=-d_{11}(b_{12}-b_{11}\frac{b_{22}}{b_{21}}),$

$C_{21}=-a_{21}\frac{c_{11}}{b_{21}},$

$C_{22}=-\left( a_{22}+a_{21}c_{12}\right)/b_{21},$

$C_{23}=-\left( b_{23}+a_{21}c_{13}\right)/b_{21},$

$C_{24}=-\left( b_{22}+a_{21}c_{14}\right)/b_{21}.$

Now, we convert Eq. (\ref{eq7}) using Eq. (\ref{eq11}) and Eq. (\ref{eq14})
\begin{multline}\label{eq16}
E=\frac{1}{2}\int_{0}^{l}{mf^{2}dz}\dot{q}^{2} +
\frac{1}{2}\int_{0}^{l}{J_{m}\phi^{2}dz}\dot{r}^{2}
-\int_{0}^{l}{m\sigma_{T}f\phi dz} \dot{q}\dot{r} 
+\frac{1}{2}\int_{0}^{l}{EJ(f'')^{2}dz}q + \frac{1}{2}\int_{0}^{l}{GJ_{k}(\phi')^{2}dz}r = \\
\frac{1}{2}a_{13}x_{1}+\frac{1}{2}a_{11}x_{2}^{2}
-\frac{1}{2}b_{23}^{(2)}x_{3}-\frac{1}{2}b_{21}x_{4}^{2}-a_{21}x_{2}x_{4}\leq E_{*}.
\end{multline}

Finally, we use integration by parts:
\begin{multline}\notag
\int_{0}^{l}{EJ(f'')^{2}dz} = \left. EJf''f' \right\rvert_{0}^{l} - 
\int_{0}^{l}{\left(EJf''\right)'f'dz} 
= 
-\frac{d\left( EJf''\right)}{dz}\left. f\right\rvert_{0}^{l} + \int_{0}^{l}{\frac{d^{2}\left( EJf''\right)}{dz^{2}}fdz} = a_{13},
\end{multline}

\begin{equation}\notag
\int_{0}^{l}{GJ_{k}\left(\phi' \right)^{2}dz} 
= GJ_{k}\phi'\left. \phi\right\rvert_{0}^{l} - \int_{0}^{l}{\frac{d\left(GJ_{k}\phi' \right)}{dz}\phi dz} 
= -b_{23}^{(2)}.
\end{equation}

Now we summurize the results of the section. We obtained an equation for the dynamics of a system that describes small flexural-twisting oscillations of a wing in a laminar flow, taking into account the linear aerodynamic force and moment, see Eq. (\ref{eq15}). Moreover, we obtained the equation for limiting the total energy of such a system, see Eq (\ref{eq16}). 
The rate of change of the angle of inclination of the 'feather' in relation to the wing plane can be selected as a control.

\section{CONTROL SYNTHESIS WITH THE SPEED-GRADIENT METHOD}

Up to this point, we have described the dynamic system taking into account how the solution should look in the optimal state, determined by Eq. (\ref{eq16}). However, we do not yet have a control law that would allow us to control the system so that it reaches this state. In this section we synthesize such a law.

First of all, we seek control for the Eq. (\ref{eq15}) with the criterion of Eq. (\ref{eq16}), using the speed-gradient method introduced in \cite{A6,A7}.

\begin{multline}\notag
\frac{dE}{dt} = \frac{1}{2}a_{13}x_{2}+a_{11}x_{2}\left( \sum_{k=1}^{4}{C_{1k}x_{k}} + F_{1}(\beta, u)\right)
-\frac{1}{2}b_{23}^{(2)}x_{4} 
- b_{21}x_{4}\left( \sum_{k=1}^{4}{C_{2k}x_{k}} + F_{2}(\beta, u)\right) \\
-a_{21}x_{4}\left( \sum_{k=1}^{4}{C_{1k}x_{k}} + F_{1}(\beta, u)\right) 
-a_{21}x_{2}\left( \sum_{k=1}^{4}{C_{2k}x_{k}} + F_{2}(\beta, u)\right),
\end{multline}

\begin{multline}\notag
\nabla_{u}\left( \frac{dE}{dt}\right) = 
col\left\{ (a_{11}x_{2}-a_{21}x_{4})\frac{\partial F_{1}}{\partial u_{i}}
-(b_{21}x_{4}+a_{21}x_{2})\frac{\partial F_{2}}{\partial u_{i}}, i={1,N}
\right\}=\\
col\left\{ (a_{11}x_{2}-a_{21}x_{4})s_{1i}
-(b_{21}x_{4}+a_{21}x_{2})s_{2i}, i={1,N}\right\}=\\
col\left\{ (a_{11}s_{1i}-a_{21}s_{2i})x_{2}
-(a_{21}s_{1i}+b_{21}s_{2i})x_{4}, i={1,N}\right\}=
col\left\{ \mu_{i}x_{2} + \nu_{i}x_{4}, i={1,N}\right\},
\end{multline}
where
$\mu_{i} = a_{11}s_{1i}-a_{21}s_{2i};~$
$\nu_{i} = -(a_{21}s_{1i}+b_{21}s_{2i}).$

Thus,
\begin{equation}\notag
\frac{du_{i}}{dt} = -\gamma_{i}(\mu_{i}x_{2}+\nu_{i}x_{4}), ~~ i={1,N}, ~~ \gamma_{i}>0
\end{equation}

or
\begin{equation}\notag
\frac{du_{i}}{dt} = -\gamma_{i}(\mu_{i}\dot{x}_{1}+\nu_{i}\dot{x}_{3}) \Rightarrow
u_{i} = -\gamma_{i}(\mu_{i}x_{1}+\nu_{i}x_{3}) + const_{i}.
\end{equation}

Since for $x_{1}=x_{2}=x_{3}=x_{4}=0$ all values of $u_{i}=0,$
and it means that $const_{i} = 0, ~~ i={1,N}.$

The resulting control equation
\begin{equation}\label{eq17}
u_{i} = -\gamma_{i}(\mu_{i}x_{1}+\nu_{i}x_{3})
\end{equation}
is a control in the form of feedback on a deviation with constant coefficients.

\section{MULTI-AGENT CONTROL}

Now, suppose that the feathers are some intelligent agents, such that each of them can receive information about the movement of the wing in the area, where it is located, exchange this information with other agents (transfer their information to them and receive their information), process the received data and form local force and moment impacts on the wing, trying to keep the wing as close to the curve as possible (see Fig. \ref{fig:2p}). Assuming the square of the feather is quite small compared to the surface of the wing, we connect the feather with some point on the wing’s surface, when it is in the neutral position, i.e. with some point on the wing surface that this feather covers.

For each feather ($i={1,N}$) we denote:
\begin{itemize}
	\item $\bar{z}_{i},\bar{\psi}_{i}$ - coordinates of the point to which the $i$-th feather is connected;
	\item $y_{1i}$ and $\Theta_{1i}$ - deflection and angle of twisting of the wing at the location of the $i$-th feather (deviations from the curve (\ref{eq2}));
	\item $N_{i}$ - the number of feathers with which the $i$-th feather can exchange information;
	\item $\bar{b}_{ij}$ - non-negative weighting coefficient of the significance of information from the $i$-th feather to the $j$-th.
Here, we assume that $\bar{b}_{ij}=\bar{b}_{ji}$
and $\sum_{j\in N_{i}}{\bar{b}_{ij}}=1$ - is the normalization condition;
	\item $\bar{b}_{ij}=0,$ if the i-th and $j$-th feathers are not connected informationally;
	\item $\bar{B}=[\bar{b}_{ij}]$ - adjacency matrix.
\end{itemize}
 
According to (\ref{eq8}) and (\ref{eq11}), for each $i={1,N}$ it is necessary to provide
\begin{equation}\notag
\begin{Vmatrix}  
w_{i}
\end{Vmatrix}
=
\begin{Vmatrix}  
y_{1i}\\ 
\dot{y}_{1i}\\
\Theta_{1i}\\
\dot{\Theta}_{1i}
\end{Vmatrix}
=
\begin{Vmatrix}  
qf(z_{i})\\ 
\dot{q}f(z_{i})\\
r\phi(z_{i})\\
\dot{r}\phi(z_{i})
\end{Vmatrix}
=
\begin{Vmatrix}  
\Phi_{i}\bar{x}
\end{Vmatrix}
< \epsilon
\end{equation}

for $t>t_{1}$ for a long period of time;
$t_{1}$ - the moment of reaching  $V_{flat}$;
$\Phi_{i}=diag\{f(z_{i}),f(z_{i}),\phi_{i}(z_{i}),\phi_{i}(z_{i})\}$

Moreover, we take into account that
\begin{equation}\notag
\left\|\Phi_{i}\bar{x} - \Phi_{j}\bar{x} \right\|
=\left\|\left[\Phi_{i} - \Phi_{j}\right]\bar{x} \right\|
\leq \left\|\Phi_{i}\bar{x}\right\| + \left\|\Phi_{j}\bar{x}\right\|
< 2\epsilon = \epsilon^{*}.
\end{equation}

We define the quality of compensation for deviations from the stationary position of the wing using the functional
\begin{equation}\label{eq18}
L(\bar{x})=\frac{1}{2}\sum_{i=1}^{N}\sum_{j\in N_{i}}{\bar{b}_{ij}\left\|\left(\Phi_{i} - \Phi_{j}\right)\bar{x} \right\|^{2}}.
\end{equation}

We formulate the problem by analogy with \cite{A5} as follows.
In conditions of uniform rectilinear flight of the aircraft in a laminar flow, when approaching at time $t_{0}$ to the critical speed of onset of (flexural-torsional) flutter wing (and further exceeding this speed), it is required to find for each feather such controls ui of the system (\ref{eq15}), which would ensure for the goal functional of Eq. (\ref{eq18}) the fulfillment of the target condition:
\begin{equation}\label{eq19}
L(\bar{x}) \leq \epsilon^{*}
\end{equation}

for a small given parameter $\epsilon_{*}>0$ for $t>t_{1}$
during a sufficiently long period of time,
where $t_{1}$ is the moment of reaching the critical flutter speed.

\subsection{NON-MULTI-AGENT CONTROL SYNTHESIS}

Let us consider the functional
\begin{equation}\notag
L(x) = \frac{1}{2}\sum_{i=1}^{N}\sum_{j\in N_{i}}{\bar{b}_{ij}\left\|(\Phi_{i}-\Phi_{j})\bar{x}\right\|}^{2}\leq \epsilon^{*}.
\end{equation}

We suppose that $\bar{b}_{ij}\geq 0~\forall i,j$ which implies $L\geq 0.$

\begin{multline}\notag
\left\|(\Phi_{i}-\Phi_{j})\bar{x}\right\|^{2} = 
\left\|diag\{a,~b, ~c,~d \}\bar{x}\right\|^{2} = 
\left\|\left(f_{ij}x_{1},~f_{ij}x_{2},~\phi_{ij}x_{3},~\phi_{ij}x_{4} \right)^{T}\right\|^{2} = \\
f_{ij}^{2}(x_{1}^{2}+x_{2}^{2}) + \phi_{ij}^{2}(x_{3}^{2}+x_{4}^{2}),
\end{multline}
where 
$a = f(z_{i})-f(z_{j})$, $b = f(z_{i})-f(z_{j})$, $c = \phi(z_{i})-\phi(z_{j})$, $d = \phi(z_{i})-\phi(z_{j})$ 
and 
$f_{ij}=f_{i}-f_{j}=f(z_{i})-f(z_{j}),$ $\phi_{ij}=\phi_{i}-\phi_{j}=\phi(z_{i})-\phi(z_{j}).$

Now:
\begin{multline}\label{eq20}
L(x) = \frac{1}{2}\sum_{i=1}^{N}\sum_{j\in N_{i}}{\bar{b}_{ij}\left[f_{ij}^{2}(x_{1}^{2}+x_{2}^{2}) + \phi_{ij}^{2}(x_{3}^{2}+x_{3}^{2}) \right]} 
= \frac{1}{2}\left[\chi(x_{1}^{2}+x_{2}^{2}) + \lambda(x_{3}^{2}+x_{4}^{2}) \right],
\end{multline}
where $\chi = \sum_{i=1}^{N}\sum_{j\in N_{i}}{\bar{b}_{ij}f_{ij}^{2}}\geq 0$ and 
$\lambda = \sum_{i=1}^{N}\sum_{j\in N_{i}}{\bar{b}_{ij}\phi_{ij}^{2}}\geq 0$
are constants, which are determined by the topology of the agent network (for given functions of the waveforms $f(z)$ and $\phi(z)$).

The case when $\chi=0$ or $\alpha=0$ is degenerate, because, in this situation, the functional (\ref{eq20}) is insensitive to the presence of feathers.

Following the Speed-Gradient method of \cite{A6} and using the equations (\ref{eq20}) and (\ref{eq15}), we obtain:
\begin{multline}\notag
\frac{dL}{dt}=\chi(x_{1}x_{2} + x_{2}\dot{x}_{2})+\lambda(x_{3}x_{4} + x_{4}\dot{x}_{4})= \\
\chi x_{2}\left[x_{1}+\sum_{k=1}^{4}{C_{1k}x_{k}} + F_{1}(\beta,u) \right] + 
\lambda x_{4}\left[x_{3}+\sum_{k=1}^{4}{C_{2k}x_{k}} + F_{2}(\beta,u) \right],
\end{multline}

\begin{equation}\notag
\nabla_{u}\dot{L} = col\left\{ \frac{\partial\dot{L}}{\partial u_{p}},~p=1,N\right\},
\end{equation}

\begin{multline}\label{eq201}
\frac{\partial}{\partial u_{p}}\left(\frac{dL}{dt}\right) = \chi x_{2}\frac{\partial F_{1}(\beta,u)}{\partial u_{p}}
+ \lambda x_{4}\frac{\partial F_{2}(\beta,u)}{\partial u_{p}} = \\
\chi x_{2}s_{1p} + \lambda x_{4}s_{2p} =
\chi s_{1p}\dot{x}_{1} + \lambda s_{2p}\dot{x}_{3}.
\end{multline}

Consequently:
\begin{multline}\label{eq21}
\frac{d u_{p}}{dt} = -\gamma_{p}\left\{ \chi s_{1p}\dot{x}_{1} + \lambda s_{2p}\dot{x}_{3}\right\} \Rightarrow \\
\dot{\beta}_{p} = u_{p} = -\gamma_{p}\left\{ \chi s_{1p}{x}_{1} + \lambda s_{2p}{x}_{3}\right\}, ~ \gamma_{p}>0,~p=1,N,
\end{multline}
since the integration constant is zero for the same reasons as in (\ref{eq17}).

The Eq. (\ref{eq17}) and Eq. (\ref{eq21}) have the same structure, but there are some differences. In fact, according to Eq. (\ref{eq15}), we can consider the coefficients s1i and s2i as the coefficients of influence of the i-th feather on the force factor in bending vibrations and on the moment factor in torsional vibrations, respectively. Actually, the values of these coefficients show the degree of participation of the i-th feather in wing dynamics.

In Eq. (\ref{eq17}), the feedback coefficients take into account the influence of the types of wing oscillations on each other, while in Eq. (\ref{eq21}) the feedback coefficients for bending and torsional vibrations are strictly separated and determined through influence factors only on its type of oscillation. Equation (\ref{eq21}) is not multiagent by nature since its dependence upon information about the state of other agents is static and it is invariant to the dynamics of the i-th feather.

\subsection{MULTI-AGENT CONTROL SYNTHESIS}
Now, we go back to Eq. (\ref{eq15}). We expand the vector of phase coordinates by introducing
\begin{equation}\label{eq22}
\tilde{x}_{i}=col\left\{ \Phi_{i}\bar{x},\beta_{i}\right\}.
\end{equation}

For this extended vector, we compose a functional – an analog of the functional of Eq. (\ref{eq20})
\begin{equation}\label{eq23}
\tilde{L}=\frac{1}{2}\sum_{i=1}^{N}\sum_{j\in N_{i}}{\bar{b}_{ij}\left\| col\left\{ (\Phi_{i}-\Phi_{j})\bar{x}, \beta_{i}-\beta_{j}\right\}\right\|^{2}}.
\end{equation}
The application of this approach can be justified by the fact that for small deviations of the wing from the stationary position, determined by Eq. (\ref{eq2}), deviations of the feathers from their neutral position $\beta_{i},~i=1,N$ should be small as well. That is, at least for feathers on one side of the wing (lower/upper), we have that
\begin{equation}\notag
\left(\beta_{i}-\beta_{j}\right)^{2}\leq \beta_{i}^{2}+\beta_{j}^{2} < 2\epsilon^{2}_{\beta},
\end{equation}
where $\epsilon_{\beta}$ is some fairly small number.

After simple transformations, we get from (\ref{eq23})
\begin{equation}\label{eq24}
\tilde{L}=L+\frac{1}{2}\sum_{i=1}^{N}\sum_{j\in N_{i}}{\bar{b}_{ij}\left(\beta_{i}-\beta_{j}\right)^{2}} < \epsilon^{*} +
\sum_{i=1}^{N}\sum_{j\in N_{i}}{\bar{b}_{ij}\epsilon^{2}_{\beta}} < \epsilon^{**}.
\end{equation}

Now, we apply the Speed-Gradient method:
\begin{equation}\notag
\frac{d\tilde{L}}{dt} = \dot{L} + \sum_{i=1}^{N}\sum_{j\in N_{i}}{\bar{b}_{ij}\left(\beta_{i}-\beta_{j}\right)\left(u_{i}-u_{j}\right)};
\end{equation}

\begin{equation}\notag
\nabla_{u}\left(\frac{d\tilde{L}}{dt}\right)=col\left\{ \frac{\partial\dot{L}}{\partial u_{p}} + 2\sum_{j\in N_{p}}{\bar{b}_{pj}(\beta_{p}-\beta_{j})}\right\}
\end{equation}

Finally, the control law takes the following form:
\begin{multline}\label{eq25}
\dot{\beta}_{p}=u_{p}=-\tilde{\gamma}_{p}\left(\chi s_{1p}\dot{x}_{1} + \lambda s_{2p}\dot{x}_{3}\right) 
-2\tilde{\gamma}_{p}\sum_{j\in N_{p}}{\bar{b}_{pj}(\beta_{p}-\beta_{j})},~ \tilde{\gamma}_{p}>0,~ p=1,N.
\end{multline}

It is important to note the multi-agent nature of the control protocol of Eq.(\ref{eq22}) and Eq. (\ref{eq25}), since the control signal for the rotation of each feather is formed on the basis of information about its own current state and on the basis of the current state of the feathers associated with it: the second term in Eq. (\ref{eq25}). At the same time, the first part 
of the law of Eq. (\ref{eq25}) describes control in the form of feedback with constant coefficients according to the speed of deviation of bending and torsional vibrations from the stationary state of Eq. (2). 

It is essential that if $\dot{x}_{1}=\dot{x}_{3}=0,$. Then this does not entail $u_{i}=0,$ since in the general case there can be $x_{1}\neq 0$ and $x_{3}\neq 0.$ 
To reduce them, it is necessary to apply a control defined by Eq (\ref{eq25}).
The expression $\dot{\beta}_{p}=u_{p}=0$ is true only in the case of complete absence of oscillations, when $x_{1} = x_{2} = x_{3} = x_{4} = 0,$ since only then $\beta_{p}=0,~p=1,N.$

The control of Eq. (\ref{eq25}) does not explicitly depend on the time, at which the critical flutter speed is reached. The last allows using this control without any changes also in the case of multiple transitions in speed across this boundary.

\section{CONCLUSIONS AND OUTLOOK}

This work is the first study of the authors, related to multiagent control of the wing with feathers aimed to avoid increasing wing oscillations when approaching flutter.
In the article
- a mathematical model of the bending-torsional vibrations of an airplane wing with controlled feathers on its surface is given;
- three different statements of the control problem are considered, which differ by the goal functional; 
- the three control laws (Eq. (\ref{eq17}), Eq. (\ref{eq21}) and Eq. (\ref{eq25})) are synthesized by the Speed-Gradient method. Only one of them: Eq. (\ref{eq25}) is multi-agent.

Eq. (\ref{eq21}) is an "intermediate" one for the synthesis of a multi-agent control law. It has a similar structure to Eq. (\ref{eq17}), but takes into account the presence of other feathers and their contribution.
However, the information about them remains static. It means that the state and dynamics of other pens is not considered. The multi-agent control law allows for each feather to take into account information about its own current state and about the current state of feathers in the area of the wing where it is located. As a rule, this allows for more precise tuning and quicker tuning to external factors, which makes this equation the most promising.

In the future, we advise to study the effectiveness of the obtained control laws and to compare them. The most critical indicator in the comparison should be the time, during which the system is able to damp vibrations to a safe level and hold them. The relevance of this indicator is due to the rather fast process of increasing wing oscillations during flutter.
Another promising area for the further research is the development of multi-agent management of feathers following the example of a swarm and research its effectiveness.

\section*{ACKNOWLEDGMENT}

Section 3 of the work was carried out within the project "Artificial Intelligence and Data Science: Theory, Technology, Industry and Interdisciplinary Research and Applications" on the state order of St. Petersburg State University.

Sections 4-5 of this work was supported by IPME RAS Russian Science Foundation (project 16-19-00057).

\end{document}